# Energy-Efficient Optical Crossbars on Chip with Multi-Layer Deposited Silicon

Hui Li[1], Sébastien Le Beux[1]*, Gabriela Nicolescu[2] and Ian O'Connor[1]

[1] Lyon Institute of Nanotechnology,
INL-UMR5270
Ecole Centrale de Lyon,
Ecully, F-69134, France

[2] Computer and Software
Engineering Dept.
Ecole Polytechnique de Montréal
Montréal (QC), Canada

\* Contact author: sebastien.le-beux@ec-lyon.fr

*Abstract* — The many cores design research community have shown high interest in optical crossbars on chip for more than a decade. Key properties of optical crossbars, namely a) contention-free data routing b) low-latency communication and c) potential for high bandwidth through the use of WDM, motivate several implementations. These implementations demonstrate very different scalability and power efficiency ability depending on three key design factors: a) the network topology, b) the considered layout and c) the insertion losses induced by the fabrication process. The emerging design technique relying on multi-layer deposited silicon allows reducing optical losses, which may lead to significant reduction of the power consumption. In this paper, multi-layer deposited silicon based crossbars are proposed and compared. The results indicate that the proposed ring-based network exhibits, on average, 22% and 51.4% improvement for worst-case and average losses respectively compared to the most power-efficient related crossbars.

## I. INTRODUCTION

Optical Network-on-Chip (ONoC) is an emerging technology considered as one of the key solutions for the future generation of on-chip interconnects. It relies on optical waveguides to carry optical signals, so as to replace electrical interconnect and provide the low latency and high bandwidth. Moreover, 3D integration technologies allow both optical and electrical layers to be stacked. Proposals for ONoC can, thus, realistically envision the integration of sufficient photonic devices for fast chip-length communications [8][2][3].

Among the proposed ONoCs, the crossbar-based solutions show good properties since they do not require any arbitration [1][4]. In such networks, the signal propagates from a source IP core to a destination IP core according to its wavelength (i.e. it relies on Wavelength Division Multiplexing, WDM), which is achieved by passive Microring Resonators (MRs). Existing optical crossbar implementations show different tradeoffs between design complexity and energy efficiency and demonstrate a relatively poor scalability. Emerging design technology relying on multi-layer deposited silicon allows efficient stacking of optical layers connected by optical vias (optical vertical coupler) [17]. This technology allows reducing the number of waveguide crossings which is a main loss source in ONoC. Using multi-layer deposited technology to improve the crossbar-based design may lead to more power-efficient network suitable for large-scale systems.

In this paper, we propose multi-layer deposited silicon based implementation of optical crossbars. We compare them with their initial single-layer implementations according to worst-case and average losses. They are key metrics to evaluate the ONoC scalability and power efficiency since the performances of all the considered optical crossbars are the same. While interconnecting from 4 to 64 IP cores, our results show that the ring-based crossbar demonstrates, on average, 36.9% and 55.2% improvements for worst-case and average losses over the related networks.

The paper is structured as follows. Section II presents the related work. Section III presents the considered architecture model. Section IV and Section V are the description of the proposed implementation of the considered crossbar. Section VI gives the results and Section VII concludes the paper.

## II. RELATED WORK

### A. Optical crossbar on chip

Optical crossbars rely on wavelength-based signal routing to realize full-connectivity among IP cores. Such ONoCs rely on passive MRs and do not require any arbitration [1][4]. Related implementations of optical crossbars lead to Matrix [16], λ-router [1], Snake [10] and ORNoC [4] networks. These networks exhibit different properties such as the number of waveguide crossings, required wavelengths, MRs, etc. Their suitability thus depends on the architecture size, the connectivity requirements and the metrics to optimize. Few studies were achieved to compare the networks for a given connectivity scenario. In [10], the authors compared λ-router, Snake and ORNoC networks for processors to memories application, showing higher energy efficiency for ORNoC. In our prior work [20], we compared the above mentioned networks for NxN architecture scenarios according to the worst-case losses in the signal paths; the results highlighted significant advantages for ORNoC. In this paper, we consider the use of multi-layer deposited silicon for the efficient implementations of optical crossbars.

### B. ONoC relying on multi-layer deposited silicon

Multi-layer deposited silicon was recently proposed to improve the implementation of ONoCs by reducing the insertion loss induced by the waveguide crossing in the same layer [6] (typically around 0.05dB [6]). Optical vias allow signals to propagate from a layer to another at the price of some losses (typically around 0.1dB [17]), which thus results in a significant reduction in the number of waveguide crossings to reduce the total loss of the whole optical path. Following this idea, MPNOC [21] was proposed to realize energy-efficient communications. In R-3PO [18], MRs located on different layers allow reconfiguring the optical bandwidth between the optical layers. In this paper, we propose multi-layer deposited silicon implementation of optical crossbars and we compare them according to the worst-case and average losses. The proposed ring-based optical crossbar, so-called ORNoC$_{ML}$, demonstrates, on average, 36.9% and 55.2% improvements of worst-case and average losses respectively compared to the other crossbars for the network scales we considered.

## III. ARCHITECTURE MODEL

### A. Architecture model overview

The considered 3D architecture is composed of an electrical layer implementing $NxN$ IP cores and two optical layers implementing ONoC, with N an even number. Figure 1 shows an instance of this architecture, where N is equal to 4. The optical network in the optical layers is composed of on-chip laser sources [9] (to remove the use of power waveguide in Corona [7], which may increase the waveguide crossings in the same layer), MRs, and photodetectors. The ONoC is connected to the IP cores with Through Silicon Vias (TSV) [15]. Numerous ONoCs relying on WDM were proposed. Among these networks, wavelength routing scheme can be used to propagate data from a source IP core to a destination IP core, thus leading to a contention-free network (without need of arbitration), with high throughput and low latency.

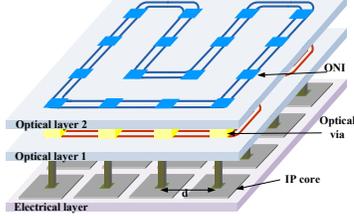

**Figure 1: The optical crossbar (ring topology in the example) is implemented in the optical layers and it interconnects IP cores**

In the interfaces of such optical crossbar, the transmitters are composed of on-chip laser sources, and the receivers are composed of photodetectors and passive MRs that drop the signal into the photodetectors (the transmitters and photodetectors are not illustrated in Figure 1). Since we consider a full connectivity between IP cores, $(N^2-1)xN^2$ laser sources, together with $(N^2-1)xN^2$ photodetectors and $(N^2-1)xN^2$ passive MRs, are required in the interface. The use of on-chip lasers is considered since, for a given wavelength, the size of an on-chip laser is of the same order of magnitude with the size of an MR used to modulate continuous waves emitted by off-chip lasers, which leads to a similar on-chip size for both approaches.

For a given communication from a source IP core to a destination IP core, the required laser output power depends on the cumulated loss along the optical path. Considering a given receiver sensitivity, the more the losses, the more the minimum laser output power, the more the ONoC power consumption. Reducing the worst-case losses is thus mandatory to reduce the overall system power consumption. The most significant losses sources are due to the propagation of signals in the waveguides, the waveguide crossing and drop. A reduction of losses can be achieved by i) improving the network topology, ii) optimizing the layout and iii) using improved fabrication process such as multi-layer silicon deposited technology.

### B. Multi-layer deposited silicon technology

Multi-layer deposited silicon technology allows efficient designs of ONoCs by stacking optical layers [6]. A key advantage of this technology is to reduce the loss of the waveguide crossing in one layer by using different layers, Figure 2-a. In this figure, red and blue colors represent photonic devices implemented in the first and second layer respectively. The design of waveguide crossing within a single layer will lead to 0.05-0.2 dB while 0 dB is expected by considering two layers. Optical vias allow transmitting signals from one layer to another. 3D photonic devices, such as MRs [6] and Photonic Switching Elements (PSEs), can also be efficiently implemented with this technology, Figure 2-b and c. However, a relatively high drop loss (e.g. 1 dB [21]) is expected for the multi-layer implementation compared to the single layer counterpart implementation (e.g. 0.5 dB [6]). Hence, a key point for the reduction of the total losses is to ensure that a "costly" drop operation from a layer to another will lead to a significant reduction in the number of waveguide crossings [21] or the waveguide length.

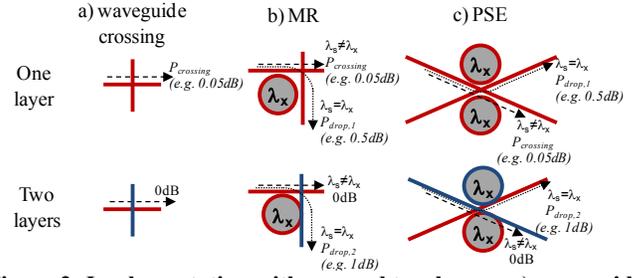

**Figure 2: Implementation with one and two layers: a) waveguide crossing, b) MR and c) PSE**

### C. Model of worst-case and average losses

The worst-case and average losses are key metrics to measure the ONoC power efficiency since lower losses lead to lower laser output power, i.e. reduced energy/bit communications. In an optical crossbar, the total loss along an optical path $L_{total}$ is given by the equation (1) and is an extension from the model proposed in [20]. $L_{total}$ depends on: the total propagation loss in the waveguide $L_{waveguide}$, the total loss due to the effective number of waveguide crossings $L_{crossing}$ (i.e. waveguide crossing occurring in a same layer, as illustrated in Figure 2-a), the drop loss in a same layer $L_{drop,1}$ (Figure 2-c), the drop loss from a layer to another $L_{drop,2}$ (Figure 2-c), and coupler loss $L_{coupler}$ introduced by optical via. The loss induced by fabrication process variation is not considered. The parameters are given in details in Table 1 and Table 2.

$$L_{total}^{dB} = L_{waveguide}^{dB} + L_{crossing}^{dB} + L_{drop,1}^{dB} + L_{drop,2}^{dB} + L_{coupler}^{dB} \quad (1)$$

$L_{waveguide} = P_{propagation} \times l_{s-d}$;
$L_{crossing} = P_{crossing} \times N_{crossing}$;
$L_{drop,1} = P_{drop,1} \times N_{drop,1}$;
$L_{drop,2} = P_{drop,2} \times N_{drop,2}$;
$L_{coupler} = P_{coupler} \times N_{coupler}$;

Each communication between IP cores is characterized according to these characteristics and, by considering a given set of insertion losses values, the total loss in the path is obtained. The worst-case loss ($L_{wc}$) and the average loss ($L_{avg}$) are thus extracted by identifying the maximum values and by applying mean function on the values. This model is generic to be used for both single-layer and multi-layer implementations, which is suitable for comparison purpose.

**Table 1 Insertion Loss Parameters**

| Parameter | Description |
|---|---|
| $P_{propagation}$ (dB/cm) | intrinsic propagation loss in a waveguide |
| $P_{crossing}$ (dB) | crossing loss |
| $P_{drop,1}$ (dB) | drop loss in the same layer |
| $P_{drop,2}$ (dB) | drop loss in $MR_{ML}$ and $PSE_{ML}$ |
| $P_{coupler}$ (dB) | vertical coupling loss (in optical vertical coupler) |

**Table 2 Network Implementation Characteristics**

| Parameter | Description |
|---|---|
| $l_{s-d}$ | waveguide length between a source and a destination |
| $N_{crossing}$ | effective number of waveguide crossings |
| $N_{drop,1}$ | number of drop operations in a same layer |
| $N_{drop,2}$ | number of drop operations from a layer to another |
| $N_{coupler}$ | number of vertical couplers in a path |

## IV. OPTICAL RINGS RELYING ON MULTIPLE LAYERS

In this section, we present ORNoC$_{ML}$, a ring-based crossbar network that can be implemented with multi-layer deposited silicon. We first present the topology and connectivity of the network and then introduce the method used for its design.

### A. Multi-layer implementation of ORNoC$_{ML}$

In ORNoC$_{ML}$, inter-IP core communication is realized through waveguides forming a ring suitable for multi-layer implementation. This is achieved without increasing the number of waveguide crossings in the same layer. The following operations are performed:

- Injection: the IP core injects an optical signal into a waveguide through its output port. The wavelength of the signal specifies the destination IP core;
- Pass through: the incoming signal propagates along the waveguide (i.e. no MR with the same resonant wavelength is located along the waveguide);
- Ejection: the incoming optical signal is ejected from the waveguide and is redirected to the destination IP core. This is achieved by one MR characterized by the same resonant wavelength with the signal.

In $ORNoC_{ML}$, the same wavelength can be used to realize multiple communications in the same waveguide at the same time. Furthermore, multiple waveguides can be used to interface IP cores and both clockwise (C) and counter-clockwise (CC) directions are considered using separate sets of waveguides. The main feature of the network is the absence of waveguide crossing, which is possible due to a) the serpentine layout and b) the use of on-chip laser sources.

Figure 3-a illustrates a structural view of $ORNoC_{ML}$ assuming 3x3 IP cores. This is presented in comparison with ORNoC illustrated in the left-hand side of the figure. The red color indicates an implementation of the waveguides in the first optical layer. Solid and dot lines are for C and CC direction, respectively. Each waveguide propagates signals from a source IP core to a destination IP core. An efficient implementation of the SWSR (Single Write Single Read) communication scheme is obtained through the reuse of a same wavelength in a same waveguide for multiple communications purpose (this is achievable through the combined use of injection and ejection for the same wavelength [4]).

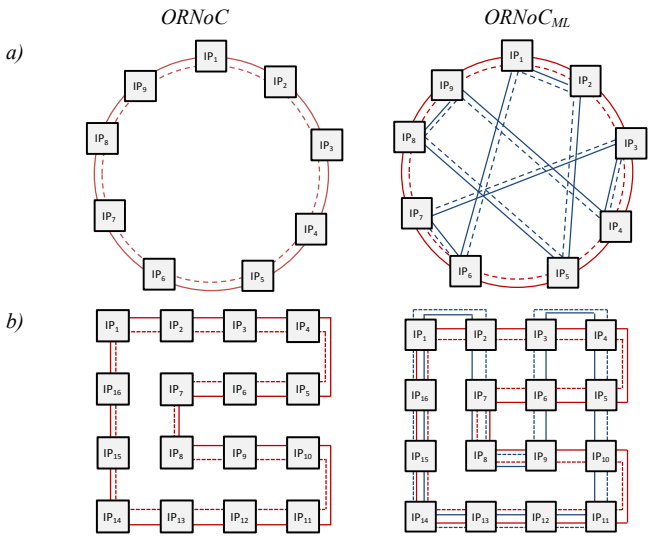

**Figure 3: Optical crossbars ORNoC and $ORNoC_{ML}$: a) topology to interconnect 9 IP cores and b) layout for 4x4 IP cores.**

### B. Ring located on multiple layers

As previously mentioned, a serpentine layout is considered to reduce waveguide crossings in the same layer. Figure 3-b shows the layout for 4x4 IP cores architecture example. Communications $IP_1 \rightarrow IP_9$ and $IP_4 \rightarrow IP_2$ in ORNoC are realized through C and CC directions respectively since they show the shortest path between the source and destination (i.e. they will demonstrate the lower propagation losses). It is worth noticing that the response paths ($IP_9 \rightarrow IP_1$ and $IP_1 \rightarrow IP_9$) will rely on opposite direction for symmetry and loss reduction purposes. $IP_1 \rightarrow IP_9$ is one of the communications that demonstrate the worst-case losses with 7 intermediate crossed interfaces. By considering a mesh distribution of the interfaces and a distance $d$ between neighborhood interfaces, the total propagation distance for this example is $8d$. This is not negligible since, on dimensional side by considering X and Y directions, connecting both IP cores with a dedicated waveguide in the same layer would require a length of $4d$ for only the waveguide layout. However, adding such a dedicated waveguide will i) introduce waveguide crossings and ii) affect the regularity, thus leading into a less scalable network.

An additional ring network is implemented in a second silicon layer, so as to reduce the propagation loss. This additional network is similar to the initial one but it is rotated from 90° in order to provide a better connectivity between the IP cores, as illustrated in the right-hand side of Figure 3-b. In this figure, red color is used to represent waveguides located in the first layer and blue one is for the second layer. Since the additional waveguides are located in a different layer, the propagation of signal does not suffer from any extra waveguide crossing in the same layer, as shown in Figure 3-b. In this example, the communications between $IP_1$ and $IP_9$ are realized in the second layer (both C and CC directions) since the propagation distance is shorter than that in the first layer. Communications between $IP_2$ and $IP_4$ are still implemented in the first layer.

It is important to notice that there is no interaction between the waveguides. Each communication between a source IP core and a destination IP core relies on only a single waveguide, either C or CC, either located in the first or second layer. In other words, signals propagating in the network are never dropped from a layer to another. For each communication, there is a single drop operation ($P_{drop,1}$) occurring to eject signals from a waveguide toward a photodetector. For the communications using the second layer, the coupling losses occurring in two optical vertical couplers are considered ($P_{coupler}$): one from layer 1 to layer 2, the second from layer 2 to layer 1. Figure 4 illustrates a layout example of an ONI. The waveguides in red and blue lines allow propagating signals in the first and second layer respectively. In this example, a single waveguide is considered for a couple of layer-direction (C-CC); however multiple waveguides can be regularly implemented without any waveguide crossing by applying the layout guidelines from [4]. For communications in the second layer, optical couplers allow interfacing the laser sources and the photodetectors with the waveguides located in the second layer.

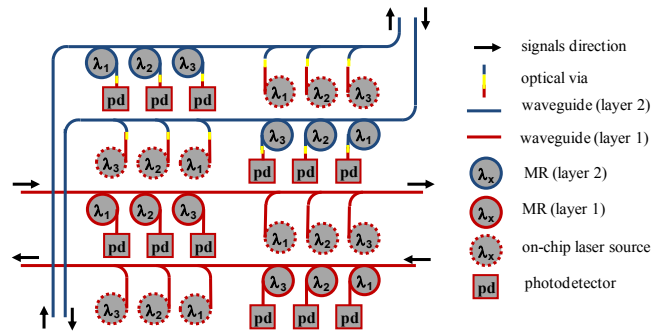

**Figure 4: An Optical Network Interface**

### C. Design method

The ring-based network is designed following a two-step methodology. First, each inter-IP core communication is assigned to the ring minimizing the propagation distance (i.e. by considering both the layer and the direction). Second, each communication is assigned to a wavelength in order to maximize the bandwidth per waveguide.

*1) Ring assignment scheme*

In $ORNoC_{ML}$, the propagation loss is reduced by implementing communications with the ring showing the shortest propagation distance between a source and a destination. For this purpose, four distance matrixes are defined, each for one possible ring (layer 1 or layer 2, C or CC direction). For each pair of source and destination, the ring with the shortest distance is assigned and the bidirectional communications (C and CC) utilize the same layer for sake of symmetry. Figure 5 illustrates an excerpt of the ring assignment matrix for the 4x4 architecture example from Figure 3. Source IP cores are represented in the first column and destination IP cores are in the first row. The communications between $IP_1$, $IP_2$, $IP_3$ and $IP_4$ are realized by using the two rings in the first layer. It is worth noticing that if equal distance is observed by using the first or the second layer (e.g. for the communication $IP_1 \rightarrow IP_2$), the first layer is assigned since it avoids the use of optical vias and the loss is less. Communication $IP_1 \rightarrow IP_9$ and $IP_9 \rightarrow IP_1$ are realized by using the second layer in C and CC directions separately. The purpose of the matrix is two folds: i) to specify the communications considered for the wavelength assignment of each ring, as detailed in the following; ii) to evaluate the worst-case and average losses in the network.

| D \ S | $IP_1$ | $IP_2$ | $IP_3$ | $IP_4$ | … | $IP_9$ | $IP_{10}$ | $IP_{11}$ | … |
|---|---|---|---|---|---|---|---|---|---|
| $IP_1$ | - | C | C | C | | C | CC | CC | |
| $IP_2$ | CC | - | C | C | | C | C | CC | |
| $IP_3$ | CC | CC | - | C | | CC | C | C | |
| $IP_4$ | CC | CC | CC | - | | CC | C | C | |
| … | | | | | | | | | |
| $IP_9$ | CC | CC | C | C | | - | C | C | |
| $IP_{10}$ | C | CC | CC | CC | | CC | - | C | |
| $IP_{11}$ | C | C | CC | CC | | CC | CC | - | |
| … | | | | | | | | | |

**Figure 5: Ring assignment matrix for 4x4 IP cores**

*2) Wavelength assignment algorithm*

The efficient design of $ORNoC_{ML}$ requires careful wavelength assignment between IP cores in order to reduce i) the number of wavelengths and ii) the number of waveguides per direction if the maximal wavelength number is reached in one waveguide.

Based on the matrix obtained from the ring assignment scheme, the wavelengths are assigned as follow: first, for each ring, starting from source core $IP_X$, a wavelength is assigned to the longest-distance communication in direction C to a destination core $IP_Y$; second, the same wavelength is assigned to the communication from $IP_Y$ to the longest-distance communication (still in direction C) to a destination core $IP_Z$. We apply the same assignment to the following longest-distance communications until source core $IP_X$ is reached, to ensure that the wavelength is used on the whole ring (i.e. the wavelength is efficiently used on the ring). The same process is applied by starting from each of other IP cores with a different wavelength. The same algorithm iterates with other wavelengths until a wavelength is assigned to each communication. If the wavelengths number reaches the maximum allowed in one waveguide, then another waveguide is added and the algorithm continues from the initial wavelength, without any impact on the layout complexity and waveguide crossing. For sake of symmetry, bidirectional communications are implemented in C and CC directions with the same wavelength.

## V. MULTI-LAYER IMPLEMENTATIONS OF RELATED OPTICAL CROSSBARS ON CHIP

For a fair comparison of the proposed network with related optical crossbars, we consider the multi-layer based implementations of crossbar networks. The considered networks are 1) Matrix [16], 2) λ-router [1] and 3) Snake [10]. For each network, we give i) a structural view example, ii) the considered layouts, iii) the required number of MRs and iv) the loss model parameters (e.g. effective number of waveguide crossings).

### A. Matrix

Figure 6-a illustrates the use of Matrix crossbar to interconnect 4 IP cores. In order to avoid any waveguide crossing in the same layer, input and output waveguides (respectively represented with red and blue lines) are located in the first and second layers respectively. This implies that the MR drops resonating signals from a waveguide located in the first layer to a waveguide in the second one (Figure 2). If full connectivity is considered, a total of 16 MRs will be required in this example. However, when considering only inter-IP core communications, the MRs located on the diagonal can be removed. For NxN IP cores, $(N^2-1) \times N^2$ passive MRs are, thus, used to implement the crossbar itself.

In order to match with the layout constraints from the regular NxN architecture, it is mandatory to link i) the transmitter part of IP cores to the input of the Matrix and ii) the output of the Matrix to the receiver part of IP cores. This is achieved by extending the network waveguides until the IP cores, assuming only X and Y directions for the waveguides. It is worth noticing that the waveguides are extended in their respective layers. Finding an optimal layout is not an easy task and it depends on various parameters such as the number of IP cores, the distance between the IP cores and the considered insertion losses. For instance, if the considered propagation losses are rather large (e.g. 2dB/cm), then a layout with waveguide crossings in the same layer may demonstrate lower losses than a layout without waveguide crossings (but with longer waveguide). Hence, for a fair comparison with the ring (which totally avoids waveguide crossings in the same layer), we assumed two layouts which are designed to a) avoid any waveguide crossing in the same layer between the network interfaces and the crossbar network itself (the network in such layout is named as $Matrix_{ML,A}$), and b) minimize the waveguide length (the network in such layout is named as $Matrix_{ML,B}$). Figure 6-b and c illustrate these layouts for the example of a 4x4 IP cores architecture. The crossbar network (i.e. matrix) is located in the middle of the optical layer for layout symmetry purposes (it is represented as a box for the sake of clarity).

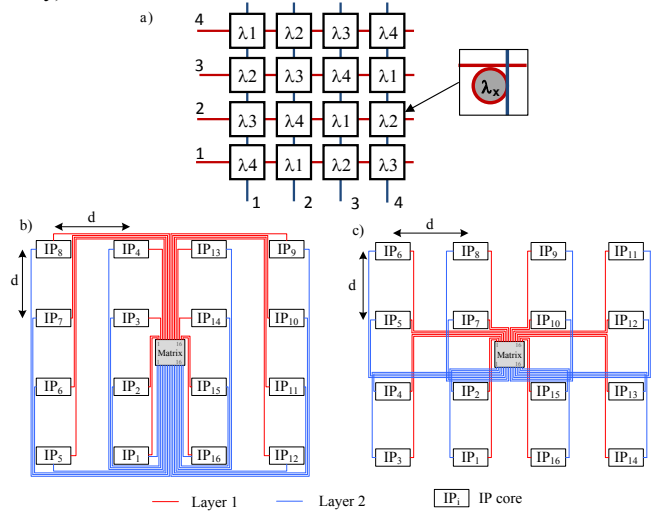

**Figure 6: a) Matrix topology, b) layout without waveguide crossings and c) layout with the shortest waveguide length**

### B. λ-router and Snake

λ-router and Snake are multistage networks sharing similar properties. The input signals propagate along waveguides and when necessary can be dropped from a waveguide to another in order to reach the targeted outputs, as illustrated in Figure 7-a and b. Compared to the Matrix, the switching structure of λ-router and Snake is a symmetric PSE implemented with 2 identical MRs. For a fair comparison with Matrix and the ring-based network, a reduction method [1] reduces the network complexity by managing only the required optical connections and removing the unused PSEs. For λ-router and Snake, the PSEs located in the central row and central column are removed respectively.

When considering a single layer for the implementation of the waveguides, the multistage topologies imply a relative large number of waveguide crossings in the worst-case path. Indeed, for networks with N inputs, there are $N^2-1$ and $2N^2-5$ waveguide crossings in the worst-case path of λ-router and Snake respectively. This can be significantly reduced by assuming two layers for the implementation of the waveguides. For sake of regularity and symmetry, the input waveguides are alternately located in the first and second layers, as illustrated in Figure 7. Following this layout design rule, the number of effective waveguide crossings (i.e. the number of waveguide crossings in the same layer) in the worst-case path drops to 12 and 13 for 4x4 λ-router and 4x4 Snake respectively, which represent a 20% and 51.9% reduction. PSEs with both waveguides located in the same layer will be implemented in a traditional way (i.e. with both MRs designed in the same layer). Otherwise, PSEs are implemented with one MR located in each layer (see Figure 2-c).

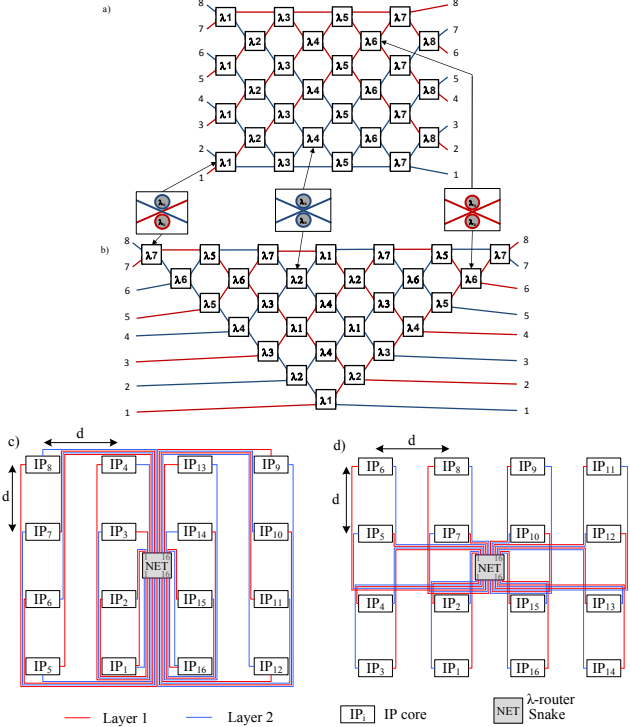

**Figure 7: a) λ-router and b) Snake and layouts c) without waveguide crossings and d) with the shortest waveguide length.**

Similarly to the Matrix, the inputs and outputs of the network are linked to the IP cores. This is achieved with the layouts illustrated in Figure 7-c and d. Since λ-router and Snake have the same interface (i.e. the same waveguides are located in layers 1 and 2), the two layouts are considered for both networks (named as λ-router$_{ML,A}$, λ-router$_{ML,B}$, Snake$_{ML,A}$, Snake$_{ML,B}$). It is worth noticing that the layout without waveguide crossings cannot be improved by considering the multiple layers for the implementation of the waveguides. However, a significant reduction in the number of waveguide crossings is achieved for the layout considering the shortest waveguide length. This will be further evaluated in the result section of the paper.

## VI. COMPARATIVE STUDY

We evaluate and compare the multi-layer based implementations according to the worst-case and average losses.

**Table 3: Insertion Loss Parameters**

| Optical loss | $P_{crossing}$ | $P_{propagation}$ | $P_{drop,1}$ | $P_{drop,2}$ | $P_{coupler}$ |
|---|---|---|---|---|---|
| Biberman [6] | 0.05 | 0.5 | 0.5 | - | 0.1 |
| Zhang [21] | 0.05 | 1 | - | 1 | - |
| Pan [11] | 0.05 | 1 | 1.5 | - | - |
| Kirman [12] | 0.12 | 1 | 1 | - | - |
| Koka [13] | 0.2 | 0.1 | 1.5 | - | - |

### A. Worst-case and average losses evaluation

Figure 8 illustrates the evaluation results of worst-case loss for the same architectures as in the previous comparison. A first observation on the worst-case loss evaluation can be made regarding the layouts: for 2x2, 4x4 and 6x6 IP cores, the layout with the shortest waveguide lengths outperforms the layout without any waveguide crossing, independently from the network topology. However, the layout without any waveguide crossing demonstrates better scalability since the loss is less impacted by the increasing number of IP cores. For 8x8 architecture, it exhibits lower losses for λ-router and Snake. Similar observation can be made for average loss (not illustrated). The results indicate that the most scalable ONoC would combine the use of i) deposited silicon to reduce waveguide crossings in the network and ii) a layout avoiding waveguide crossing between the network and IP cores. These criteria are gathered into ORNoC$_{ML}$, which demonstrates the lowest worst-case loss despite the long distance introduced by the serpentine layout. For 8x8 case, the worst-case path in ORNoC$_{ML}$ is 4.5dB, followed by Matrix$_{ML,B}$ with 4.75dB. By considering the average loss, ORNoC$_{ML}$ allows reducing the average loss by 55.2% on average compared to the other implementations. This significant difference is obtained through the shorter propagation distance between neighborhood IP cores. The improvement reaches 54% for 2x2 case and it increases to 56.3% for 8x8 case, where the average loss is 2.02dB for ORNoC$_{ML}$ compared to 4dB for Matrix$_{ML,B}$. By considering the use of tunable lasers output power [3], ORNoC$_{ML}$ will thus demonstrate significant power saving compared to the other crossbar implementations.

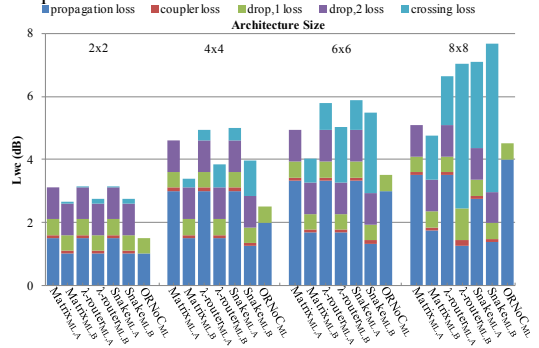

**Figure 8: Worst-case losses evaluation for 2x2 to 8x8 IP cores by using parameters from Biberman [6]**

Figure 9 represents the comparison results for a fixed number of IP cores (6x6) and various distances between them (d=1, 1.5, 2, 2.5 and 3mm). The increase of the loss with the distance is higher for the networks relying on the layout without any waveguide crossing. In any case, and despite the serpentine layout, even for the longest distance we considered (3mm, which leads to a realistic 3.24cm² die size), ORNoC$_{ML}$ is the most power-efficient network.

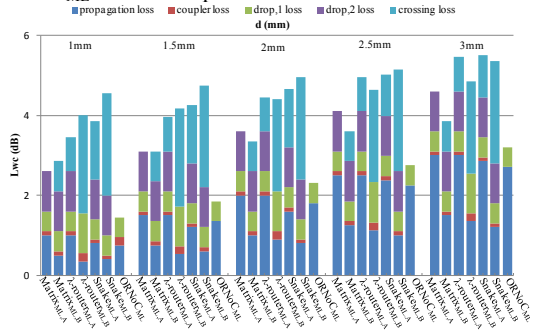

**Figure 9: Crossbar comparison for 6x6 IP cores in worst-case loss with distance between IP cores ranging from 1mm to 3mm by using parameters from Biberman [6]**

For a 8x8 architecture, the implementation of Matrix requires 63 wavelengths wrt 64 wavelengths for Snake and λ-router. Such value is rather high with regard to the high level of crosstalk and it will lead to the fabrication variability. A more reasonable implementation would be to consider several smaller networks, which implies additional waveguide crossings [14]. The use of the ring topology intrinsically leverages this issue since the number of waveguides can be set according to the crosstalk and process variability requirements. This can be achieved without any waveguide crossing because of the multi-layer implementation and the use of on-chip laser sources. Following the methodology from [4], the ring would require 16 waveguides if we consider the optimistic maximum number of 64 wavelengths per waveguides, and 63 waveguides if we consider more realistic scenario with 16 wavelengths per waveguide. Regarding $ORNoC_{ML}$, no waveguide crossing is required and the layout is regular, which makes the network scalable without any custom place-and-route tool [10][5].

*B. Comparison through insertion loss ranges values*

The comparisons from the previous sections are achieved with a given set of insertion losses values, which may lead to incomplete and/or unfair comparisons according to the implementation of the network. For instance, considering low propagation losses and high crossing losses values will give an advantage to the layout without any waveguide crossing over the layout with the shortest waveguide length and also to $ORNoC_{ML}$ over the other crossbars. For this purpose, we further compare $ORNoC_{ML}$ and $Matrix_{ML,B}$ (i.e. showing the closest losses values compared to the ring) crossbars for 8x8 scale by considering a range of 0-2dB/cm for propagation losses and a range of 0-0.2dB for waveguide crossing loss. Figure 10 illustrates the comparison results on the worst-case loss for different distances between IP cores (i.e. 1, 1.5, 2 and 2.5mm). For a given distance, we evaluate for which propagation loss value $Matrix_{ML,B}$ approximates $ORNoC_{ML}$. The area below a line represents the design space for which $ORNoC_{ML}$ provides lower worst-case losses; the area above the line gives the design space where the worst-case loss is lower for $Matrix_{ML,B}$, and the line itself represents the designs with the same worst-case losses for both implementations. For instance, $ORNoC_{ML}$ exhibits lower worst-case loss for all the insertion losses values from Table 3 for a distance between IP cores below or equal to 1mm. For 2.5mm, $Matrix_{ML,B}$ outperforms $ORNoC_{ML}$ when values from Pan [11] are considered and both network implementations show equivalent worst-case losses with the value from Kirman [12]. Overall, $ORNoC_{ML}$ performs better especially when the distance becomes smaller, even compared to the $Matrix_{ML,B}$ which outperforms other architectures in worst-case losses. Regarding the average losses (not illustrated), $ORNoC_{ML}$ exhibits lower loss than $Matrix_{ML,B}$ for all the insertion loss values from Table 3 and all the considered distances.

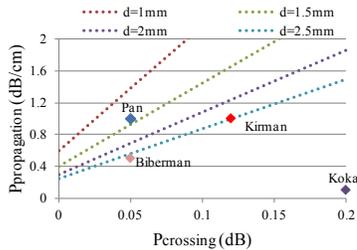

**Figure 10: Comparison between $Matrix_{ML,B}$ and $ORNoC_{ML}$ (8x8 IP cores, $P_{drop,1}$=0.5dB, $P_{drop,2}$=1dB)**

*C. Summary of the main results and discussion*

$ORNoC_{ML}$ provides the lowest worst-case loss for 2x2 to 8x8 architectures over Matrix, λ-router and Snake networks. For realistic chip sizes (4 cm$^2$), $ORNoC_{ML}$ performs 22% and 51.4% improvement for worst-case and average losses compared to $Matrix_{ML,B}$. In any cases, $ORNoC_{ML}$ outperforms the related crossbar implementations regarding the average loss metric. For larger architecture size, e.g. 10x10, Matrix, Snake and λ-router will reach a physical limitation related to the maximum number of wavelengths per waveguide (i.e. 64 wavelengths [19]). A solution to tackle this limitation is to replicate the network implementation, which leads to additional waveguide crossings and less regular layout. $ORNoC_{ML}$ demonstrates a better scalability since the use of extra waveguides allows keeping the layout regular without any waveguide crossing.

## VII. CONCLUSION

In this paper, we proposed implementation of optical crossbar on chip utilizing multi-layer deposited silicon design technology. The ring based topology demonstrates an average of 36.9% and 55.2% improvement for worst-case and average losses respectively compared to multi-stage and matrix based topologies. We demonstrate the superiority of ring-based optical networks despite the use of technology allowing drastic reductions in the number of waveguide crossings.

## ACKNOWLEDGMENT

Hui LI is supported by China Scholarship Council (CSC).